\documentclass[aps,prd,preprint,nofootinbib]{revtex4}
\usepackage{slashed}
\usepackage{amsfonts}
\usepackage{multirow}
\usepackage{mathrsfs}
\usepackage{graphicx}
\usepackage{amsmath}
\usepackage{amssymb}
\usepackage{bm}
\usepackage{bbm}
\usepackage{color}
\def\OMIT#1{}

\newcommand{\beq}{\begin{equation}}
\newcommand{\eeq}{\end{equation}}
\newcommand{\bqa}{\begin{eqnarray}}
\newcommand{\eqa}{\end{eqnarray}}

\begin{document}

\title{\large\bf NNLO QCD Corrections to $\gamma + \eta_c(\eta_b)$ Exclusive Production in Electron-Positron Collision}

\author{Long-Bin Chen$^{1}$\footnote{chenlongbin10@mails.ucas.ac.cn}}
\author{Yi Liang$^{2}$\footnote{alavan.yi@foxmail.com}}
\author{Cong-Feng Qiao$^{2,3}$\footnote{qiaocf@ucas.ac.cn, corresponding author}}

\affiliation{$^1$School of Physics \& Electronic Engineering, Guangzhou University, Guangzhou 510006, China\vspace{5pt}\\
$^2$School of Physics, University of Chinese Academy of Sciences, YuQuan Road 19A, Beijing 100049, China\vspace{5pt}\\
$^3$Department of Physics \& Astronomy, York University, Toronto, ON M3J 1P3, Canada}

\author{~\\}

\begin{abstract}
Based on the NRQCD factorization formalism, we calculate the next-to-next-to-leading order QCD corrections to the heavy quarkonium $\eta_c (\eta_b)$ production associated with a photon at electron-positron colliders. By matching the amplitudes calculated in full QCD theory to a series of operators in NRQCD, the short-distance coefficients up to NNLO QCD radiative corrections are determined. It turns out that the full set of master integrals that we obtained could be analytically expressed in terms of Goncharov Polylogarithms, Chen's iterated integrals, and elliptic functions, which mostly do not exist in the literature and could be employed in the analyses of other physical processes. In phenomenology, numerical calculations of NNLO K-factors and cross sections of $e^+e^-\rightarrow \gamma + \eta_c(\eta_b)$ processes in BESIII and B-factory experiments are performed, which may stand as a test of the NRQCD higher order calculation while confronting to the data.

\end{abstract}

\pacs{\it 12.38.Bx, 13.60.Le, 14.40.Pq}

\maketitle

The study of heavy quarkonium production and decay is very important in the understanding of Quantum Chromodynamics (QCD), and heavy quarkonium as well. In the analysis of hadron, to overcome the obstacle of nonperturbative QCD is inevitable. Fortunately, the advent of Nonrelativistic Quantum Chromodynamics (NRQCD) factorization formalism enables the calculation of quarkonium production and decay in a solid footing \cite{NRQCD}, rather than models. NRQCD can systematically factorize the nonperturbative sectors out and leave others perturbative QCD(pQCD) calculable. Up to now NRQCD achieves a great success in  Born level and first order pQCD calculations \cite{n.brambilla}, though some unsatisfactory still remains in phenomenological study.

Since the energy scales set by heavy quarks, the charm and bottom quarks, are moderate, the strong interaction in this regime is perturbative expandable, whereas does not converge quickly. The large discrepancy between leading order theoretical calculation \cite{belle-leading-order} and experimental data \cite{abe:2002,aubert:2005}, for example in double charmonium production at B-factory, can be indeed amended by the next-to-leading order(NLO) QCD corrections \cite{Zhang:2005cha,Zhang:2006ay}. Higher order calculation itself is critical, challenging and meaningful theoretically, in addition to its phenomenological impact.

The processes of $\eta_c(\eta_b)$ exclusive production in accompany with a photon at electron-position colliders is right now an interesting topic, because it can be measured accurately and calculated precisely. Furthermore, naive estimation indicates that $e^+e^- \rightarrow \gamma+\eta_c$ cross section overshoots that of $e^+e^- \rightarrow J/\psi+\eta_c$ process by an order of magnitude, and hence $e^+e^- \rightarrow \gamma +\eta_c(\eta_b)$ processes may be well measured in Belle II experiment soon. On theory side, the leading-order calculation and next-to-leading order corrections are ready \cite{shifman,Sang:2009jc,Li:2009ki}, where the radiative corrections are negative.

In calculating high order corrections, the central issue is to calculate the multi-loop Feynman integrals, namely master integrals (MI). The first complete NNLO analytical calculations for quarkonium production and decay, the  $J/\psi(\Upsilon) \rightarrow e^+ e^-$ and $e^+ e^-\rightarrow J/\psi(\Upsilon)$ processes, were achieved in Refs.\cite{beneke:1998} and \cite{czarnecki:1998} respectively, and a few years ago even the NNNLO result for $\Upsilon \rightarrow e^+ e^-$ process was obtained \cite{Beneke:2014qea}. In recent years, at the NNLO level many calculations are performed for quarkonium production and decays \cite{czarnecki:2001, bc22, Chen:2015csa}. It is interesting to note that in recently, the NNLO corrections to $\gamma\gamma^* \rightarrow \eta_c(\eta_b)$ transition form factor and $\eta_c(\eta_b) \rightarrow light\  hardrons$ were calculated numerically \cite{Feng:2015uha,Feng:2017}. Numerical calculation is an unique and promising way for higher order radiative corrections, nevertheless right now it  experiences the shortage of proper numerical packages, especially for the kinematics in physical region.

In this work, we are going to calculate analytically the next-to-next-to-leading order(NNLO) QCD corrections to exclusive $\eta_c (\eta_b)$ production associated with a photon in electron-positron collision, the multi-scale $e^+e^- \rightarrow \gamma +\eta_c(\eta_b)$ processes. Numerical evaluation for BESIII and Belle II experiments will be presented. Note, since the kinematics of the master integrals in our calculation lie obviously in the physical region, due to the lack of corresponding numerical calculation package, to our knowledge a full numerical calculation of those integrals is still unrealistic and unreliable.

\begin{figure}[t]
\begin{center}
\includegraphics[scale=0.5]{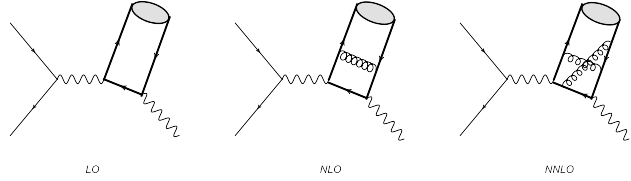}
\caption{Typical LO, NLO, and NNLO Feynman diagrams of the $e^+e^-\rightarrow \gamma+\eta_c(\eta_b)$ processes.}
\label{sample}
\end{center}
\end{figure}

In our calculation, {\bf FeynArts} \cite{feynarts} is employed to generate the corresponding Feynman diagrams and amplitudes up to NNLO, among them both massive and massless ``light by light" diagrams are taken into account. Typical LO, NLO, and NNLO Feynman diagrams of the $e^+e^-\rightarrow \gamma+\eta_c(\eta_b)$ processes are shown in Figure \ref{sample}. We calculate the amplitude and reduce the loop integrals in full QCD, before applying the NRQCD projector to quarkonium stte. Up to any loop, the fermion chains in the amplitude of concerned process can be generally expressed as
\bqa
\mathcal{M}^{\mu1\mu2}=\bar{u}(k_q)\cdot\gamma^{\mu1}\cdot\gamma^{\mu2}\cdot\slashed{k}_{\gamma}\cdot v(k_q)F_1+\ldots.
\label{amp0}
\eqa
Here, $F_1$ is a function of kinematics, coupling, renormalization and factorization scales. In above expression, the leptonic current is implied being factorized out, and the ellipses represent terms with the number of $\gamma$ matrix less than 3. Of the $e^+e^- \rightarrow \gamma^* \rightarrow \gamma+\eta_c(\eta_b)$ processes, only the first term in (1) contributes. In order to get $F_1$, we multiply (\ref{amp0}) with the projector
\bqa
P_{\mu1 \mu2} &=& [\bar{v}(k_q)\cdot\slashed{k}_{\gamma}\cdot\gamma_{\mu2}\cdot\gamma_{\mu1}\cdot u(k_q)
-\bar{v}(k_q)\cdot\gamma_{\mu1}\cdot\gamma_{\mu2}\cdot\slashed{k}_{\gamma}\cdot u(k_q)]/\nonumber\\
& &[(D-3)(D-2)(s-4m_q^2)^2],
\eqa
and sum over the spinor helicities and trace over the dirac chains. Note that by means of this procedure, one can avoid the ambiguity in the definition of heavy quarkonium projector and $\gamma_5$ in the dimensional regularization scheme.

\begin{figure}[tbh]
\begin{center}
\includegraphics[scale=0.45]{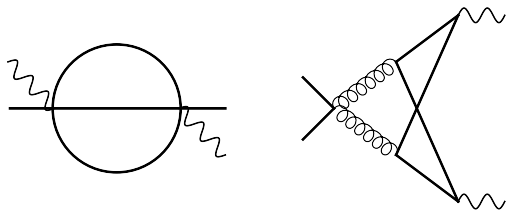}
\caption{Typical master integrals in elliptic sectors.}
\label{mi}
\end{center}
\end{figure}

With the reduction of the obtained two-loop amplitudes, we find all related scalar integrals may be attributed to a set of 133 master integrals. Using the method of differential equation \cite{Kotikov:1990kg, Kotikov:1991pm}, and by virtue of the recent development in choosing basis properly \cite{Henn:2013pwa},  it was found that 86 of the 133 master integrals can be expressed as functions of Harmonic and Goncharov polylogarithms \cite{Chen:2017xqd}. The calculation of massive multi-loop Feynman integrals is a challenging task due to the appearance of new mathematical structures, which cannot be reduced to the well-known multiple polylogarithms yet. Thanks to the very recent developments in differential equation technique in dealing with multiscale Feynman integrals \cite{Adams:2017tga,Remiddi:2016gno,vonManteuffel:2017hms,Bonciani:2016qxi}, we are now able to calculate those massive multi-loop Feynman integrals of our concern beyond multiple polylogarithms, which are found can be classified into two categories, the elliptic sectors. One elliptic sector contains the two-loop sunrise integrals with full massive propagators and those integrals containing them as subtopologies, the other sector comprises the non-planar two-loop three-point integrals. The typical Feynman diagrams of those two elliptic sectors are shown in Figure \ref{mi}.

The problems of massive two-loop sunrise integrals have been solved in Ref.\cite{Remiddi:2016gno}, where the results are expressed in terms of one-fold integrals over complete elliptic integrals and polylogarithms. We find that with the basis chosen for sunrise integrals in Ref.\cite{Remiddi:2016gno} and the method for properly choosing basis proposed in Ref.\cite{Henn:2013pwa}, differential equations for all integrals containing massive sunrise integrals as sub-topologies can be reduced to a compact form, different from \cite{Henn:2013pwa}, that are ready to be solved recursively. In our calculation, we solve them order by order up to weight four, the maximum for two-loop integrals.
For the non-planar two-loop three-point integrals, we find that the corresponding homogenous second order differential equations can be transformed linearly to the exact form of the second order differential equation for complete elliptic integrals. Then we can readily obtain the homogeneous solutions and the full solution by solving the conventional second order differential equations. Details about the solutions for elliptic sectors will be presented elsewhere.

By taking the above procedures, all the 133 master integrals can be analytically expressed as Goncharov Polylogarithms, Chen's iterated integrals, and integrals over elliptic functions and polylogarithms. Note, in elliptic sectors two-fold iterated integrals are employed to express the integrals. In order to ensure our analytical results being correct, all of them have been checked against the numerical package yields \cite{Smirnov:2015mct,Borowka:2015mxa} in both physical and non-physical regions. For each integral encountered, we have used at leat two different numerical methods to check.

About renormalization, for quark wave function and mass, $Z_2$  and $Z_m$ respectively, we adopt the on-shell scheme \cite{l2z2,z22}, while for the strong coupling constant the $\overline{\text{MS}}$ scheme is taken up to one-loop order. All ultraviolet divergences are then removed after taking the renormalization procedure. The remaining infra divergences are process independent, and can be canceled by the corrections of quarkonium NRQCD matrix element under $\overline{\text{MS}}$ scheme, with the results depending on NRQCD factorization scale $\ln\Lambda_\mu$.

Up to NNLO, the cross section of $e^+e^- \rightarrow \gamma^*\rightarrow \gamma+\eta_c(\eta_b)$ processes can be formulated as
\beq
\sigma=\sigma^0(1+\frac{\alpha_s}{\pi}c_1+(\frac{\alpha_s}{\pi})^2c_2)\ .
\eeq
Here, the LO cross section is well-known and can be reproduced easily, i.e.,
\beq
\sigma^0=\frac{32e_Q^4\pi^2\alpha^3}{3m_Q s^2} (1-\frac{4 m_Q^2}{s})\langle\mathcal{O}_1(^1S_0)\rangle\ ,
\eeq
with $\langle\mathcal{O}_1(^1S_0)\rangle$ being the NRQCD matrix element. The NLO correction coefficient $c_1$ was obtained in Refs. \cite{Sang:2009jc,Li:2009ki}, and our calculation agree with them. It reads,
\begin{eqnarray}
c_1 &=& -\frac{y+2}{2y}\text{Li}_{2}(y+1)-\frac{y+3}{4y} \ln^2 (\frac{\sqrt{y}-\sqrt{y+2}}{\sqrt{y}+\sqrt{y+2}})- \frac{(y+2)(3y+2) \ln(-y)}{4(1+y)^2}\nonumber\\
& & -\frac{3\sqrt{y+2}\ln(\frac{\sqrt{y}-\sqrt{y+2}} {\sqrt{y}+\sqrt{y+2}})}{2\sqrt{y}}-\frac{\pi^2(y+1)(2y+7)+3y(9y+8)}{12y(y+1)}\ ,
\end{eqnarray}
with $y=\frac{s-4 m_Q^2}{2 m_Q^2}$. The NNLO correction coefficient $c_2$ obtained in this work is a function of $(s,m_Q,\mu_r,\mu_\Lambda)$, which can be formulated as
\beq
c_2 = \frac{\beta_0}{4} \ln(\frac{\mu_r^2}{m_Q^2})C_F c_1 - \pi^2(C_F^2 + \frac{C_A C_F}{2}) \ln(\frac{\mu_\Lambda}{m_Q}) + f(\frac{s}{m_Q^2})\ ,
\eeq
in which the color factor $C_F = \frac{4}{3}$, $s$ represents the center-of-mass-system(CMS) energy squared, $m_Q$ denotes the heavy quark mass, $\mu_r$ is the renormalization scale, and $\mu_\Lambda$ for the factorization scale. The function $f(\frac{s}{m_Q^2})$ is quite lengthy, containing multiple polylogarithms, iterative integrals and elliptic integrals, which is not suitable to show in a journal paper, but may be provided in electronic form upon request. To guarantee our result being reliable, we have also reproduced the NNLO correction result for $\eta_c\rightarrow \gamma\gamma $ process \cite{czarnecki:2001}.

Before performing numerical calculation, we need first to fix the input parameters. The number of active quarks $n_f$ is taken to be 4 for $\eta_c$ production, and 5 for $\eta_b$ process. The charm quark mass lies in the region of $1.4$ to $1.5$ GeV, and the bottom quark mass in $4.7$ to $4.8$ GeV, to account for the heavy quark mass dependence of final result. Of the strong coupling, we first choose the well-determined value at Z-pole, that is  $\alpha_s(M_Z)=0.1184$ \cite{pdg}, and then evolve it to the heavy quark mass scales by employing the program \textbf{RunDec} in four-loop accuracy \cite{Chetyrkin:2000yt}. The NRQCD matrix elements for heavy quarkonium are taken from fittings in Refs. \cite{Bodwin:2007fz,Chung:2010vz}, i.e.,
\bqa
\langle\mathcal{O}_1(^1S_0)\rangle_{\eta_c} = 0.470\ \text{GeV}^3\ ,\ \
\langle\mathcal{O}_1(^1S_0)\rangle_{\eta_b} = 3.069\ \text{GeV}^3 \ .
\eqa

For $\gamma+\eta_c(1S)$ exclusive production in BESIII experiment, the renormalization scale $\mu_r$ is set to be at $2m_Q$. It was found in Refs. \cite{beneke:1998,Feng:2015uha} that the choice of factorization scale $\mu_\Lambda=1$ GeV results in a better convergence in perturbative expansion rather than the conventional choice of $\mu_\Lambda=m_c$ for charmonium. We find this conclusion still holds for $\gamma+\eta_c$ exclusive production, and hence $\mu_\Lambda=1$ GeV is taken in our numerical evaluation. While for bottomonium, the NRQCD factorization appears to work reasonably well, and the conventional choice of $\mu_\Lambda=m_Q$ is adequate to get a good convergence in pertubative expansion.

\begin{figure}[tbh]
\begin{center}
\includegraphics[scale=0.32]{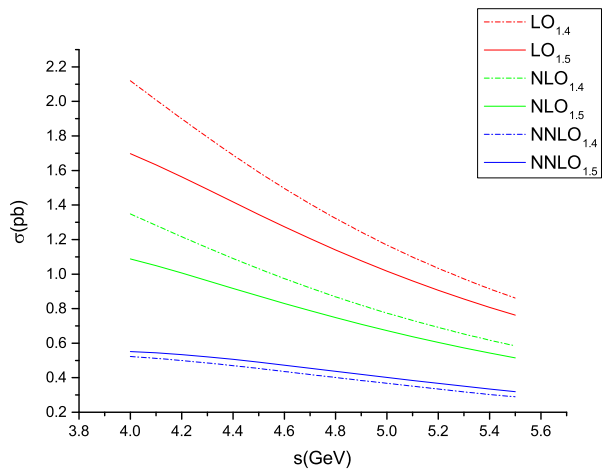}
\caption{The cross sections of exclusive $\gamma+\eta_c$ production up to NNLO in electron-positron collision with center-of-mass energy from $4.0$ to $5.5$ GeV. The subscript denotes the magnitude of charm quark mass.}
\label{svary}
\end{center}
\end{figure}

In Figure \ref{svary} we show respectively the LO, NLO, and NNLO cross sections versus the CMS energy, which varies in between $4.0$ to $5.5$ GeV.  From the figure we can read that NLO and NNLO corrections are both negative. With the NNLO correction, the production rate in this region is greatly suppressed, about 3 times reduced from the LO result, and the NNLO cross section varies more smoothly than LO and NLO results with the change of CMS energy, as expected. Recently, BESIII collaboration performed a measurement on $e^+e^-\rightarrow \gamma+\eta_c$ process with CMS energy between 4.01 and 4.60 GeV in data set corresponding to a total integrated luminosity of 4.6 $\text{fb}^{-1}$ \cite{Ablikim:2017ove}, and found no evidence of direct $\gamma + \eta_c(1S)$ production, which is in accord with the higher order QCD predictions. That is to say, the NNLO result also prefers the enhancement in $e^+e^-\rightarrow \gamma+\eta_c$ process between 4.23 and 4.36 GeV being from exotic particle $Y(4260)$ decay.

Quarkonium study in B-factory has an unique importance, where some critical measurement had ever been achieved. With the SuperKEKB/BelleII run in near future, it is reasonable to expect more from it on quarkonium physics, including the measurement on $\gamma+\eta_c(\eta_b)$ exclusive production. In Table \ref{table1} the cross sections of $\gamma+\eta_c(\eta_b)$ production at B-factory energy $\sqrt{s}=10.6~\text{GeV}$ are presented up to NNLO corrections. With the target luminosity of $8\times 10^{35}\ \text{cm}^{-2}\text{s}^{-1}$ of SuperKEKB, the exclusive $e^+e^-\rightarrow \gamma + \eta_c$ and $e^+e^-\rightarrow \gamma + \eta_b$ processes should be observed, at least the former, otherwise there will be a surprise. The NLO and NNLO corrections at B-factory are still negative, and the NNLO cross sections are some half of the LO predictions. Note that though NNLO corrections for  $e^+e^- \rightarrow \gamma + \eta_c$ is highly suppressed, it still surpasses the NLO cross section of $e^+e^- \rightarrow J/\psi + \eta_c$ process. Therefore, the measurement of exclusive processes $e^+e^-\rightarrow\gamma+\eta_c$ and $e^+e^-\rightarrow \eta_c + l^+ + l^-$, $l$ denoting leptons, are highly desired to further examine the quarkonium production mechanism and high-order QCD calculation reliability.
\begin{table}
\begin {center}
\caption{Cross sections of $e^+e^-\longrightarrow \gamma+\eta_c(\eta_b)$ processes up to NNLO accuracy at the B-factory, i.e, $\sqrt{s}=10.6$ GeV. The renormalization scale here is set to be $\mu_r=\sqrt{s}/2$.}
\vspace{12pt}
\begin{tabular}{|c|c|c|c|}
\hline
$~\sigma$(fb)~ & ~~~LO~~~ & ~~NLO~~ & NNLO  \\ \hline
$\eta_c$(1.4) & 89.7  & 75.2  &  44.6  \\ \hline
$\eta_c$(1.5) & 82.8  & 68.5  &  45.2  \\ \hline
$\eta_b$(4.7) & 2.50  & 1.77  &  1.75  \\ \hline
$\eta_b$(4.8) &  2.07 & 1.47  &  1.46  \\ \hline
\end{tabular}
\label{table1}
\end {center}
\end{table}
\begin{figure}[tbh]
\begin{center}
\includegraphics[scale=0.33]{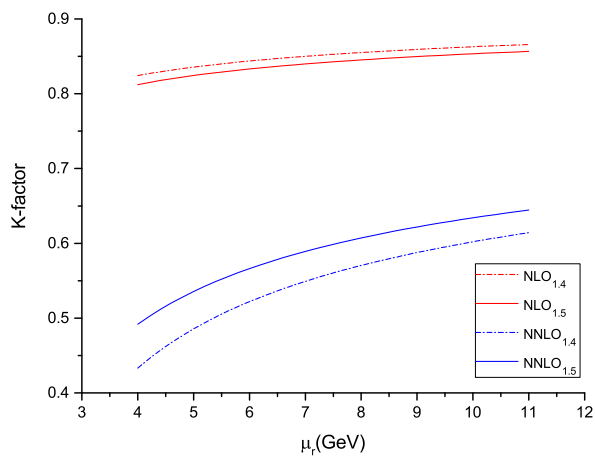}
\caption{The renormalzation scale dependence of NLO and NNLO K-factors for $\gamma+\eta_c$ exclusive production at the  B-factory. Here, the factorization scale $\mu_\Lambda= 1$ GeV, subscripts $1.4$ and $1.5$ mean the charm quark mass in GeV.}
\label{kfacc}
\end{center}
\end{figure}

\begin{figure}[thb]
\begin{center}
\includegraphics[scale=0.33]{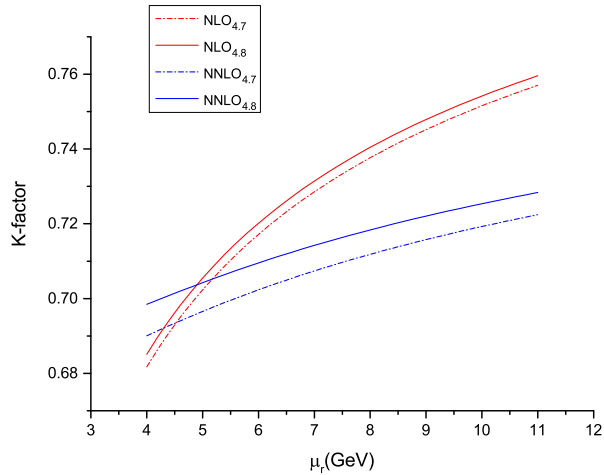}
\caption{The renormalzation scale dependence of NLO and NNLO K-factors for $\gamma+\eta_b$ exclusive production at the  B-factory. Here, the factorization scale $\mu_\Lambda= m_b$, subscripts $4.7$ and $4.8$ mean the bottom quark mass in GeV.}
\label{kfacb}
\end{center}
\end{figure}

In Figs. \ref{kfacc} and \ref{kfacb} we show the renormalization scale dependence of NLO and NNLO K-factors in $\gamma+\eta_c$ and $\gamma+\eta_b$ production at B-factory respectively. It is noticeable that though the NLO and NNLO K-factors are all negative, for $\gamma+\eta_c$ production the NNLO corrections are large and give no help to diminish the final result renormalization scale dependence, even with the choice of factorization scale at $\mu_\Lambda=1$ GeV. Whereas for $\gamma+\eta_b$ production, the NNLO corrections are moderate, the NNLO corrections reduce the renormalization scale dependence evidently. These observations imply that the perturbative expansion converges well for $\gamma+\eta_b$ production, but not for its counterpart in charm sector, which casts a shadow over the higher order corrections and further suggests that NRQCD factorization works more soundly for bottomonium sytem.

In summary, we calculate the NNLO  QCD corrections for $\gamma+\eta_c(\eta_b)$ exclusive production in electron-positron collision within the NRQCD formalism. The result tells that the NRQCD factorization still works well at the two-loop level. Thanks to the recently development in differential equation method, we find that all the master integrals can be analytically expressed in terms of Goncharov Polylogarithms, Chen's iterated integrals, and elliptic functions, which were mostly given in Ref.\cite{Chen:2017xqd}. To the best of our knowledge, this is the first time that a NNLO calculation for multiscale process about heavy quarkonium has been achieved analytically. Numerical results indicate that the NLO and NNLO cross sections are both negative in charmonium and bottomonium production. In BESIII regime, the cross section is greatly suppressed with NNLO QCD correction, which agrees with the experimental measurement and implies the $Y(4260)\rightarrow \gamma+\eta_c(1S)$ decay mode within data. Hopefully, in the forthcoming SuperKEKB run and Belle II measurement on $\gamma+\eta_c$ and $\gamma+\eta_b$ production, quarkonium production mechanism might be further elucidated.

\begin{acknowledgments}
\noindent This work was supported in part by the Ministry of Science and Technology of the Peoples' Republic of China(2015CB856703); by the Strategic Priority Research Program of the Chinese Academy of Sciences(XDB23030100); and by the National Natural Science Foundation of China(NSFC) under the grants 11375200 and 11635009.
\end{acknowledgments}

\end{document}